\documentclass[letter,twocolumn]{jpsj3}

\usepackage{times}
\usepackage{graphicx}

\def\runauthor{Takashi {\sc Hotta}}

\title{Weak-Coupling Theory for Multiband Superconductivity
Induced by Jahn-Teller Phonons}

\author{Takashi {\sc Hotta}}

\inst{Department of Physics, Tokyo Metropolitan University,
1-1 Minami-Osawa, Hachioji, Tokyo 192-0397, Japan}

\recdate{\today}

\abst{
Emergence of superconductivity in a two-band system
coupled with breathing and Jahn-Teller phonons is
discussed in a weak-coupling limit.
With the use of a standard quantum mechanical procedure,
the phonon-mediated attraction is derived.
From the analysis of the model including such attraction,
a BCS-like formula for a superconducting
transition temperature $T_{\rm c}$ is obtained.
When only the breathing phonon is considered,
$T_{\rm c}$ is the same as that of the one-band model.
On the other hand, when Jahn-Teller phonons are active,
$T_{\rm c}$ is significantly enhanced by the interband attraction
even within the weak-coupling limit.
Relevance of the present result to actual materials
such as iron pnictides is briefly commented.
}

\kword{Jahn-Teller phonon, multiband, superconductivity}

\begin{document}
\maketitle


After the Bardeen-Cooper-Schrieffer (BCS) theory
for superconductivity,\cite{BCS}
it has been pointed out that anisotropic Cooper pairs can be
formed by the Friedel oscillation in electron systems,
mainly from an academic viewpoint.\cite{Luttinger}
Nowadays it is widely recognized that anisotropic
superconductivity originating from
strong electron correlation confirms
an important route to achieve high superconducting
transition temperature $T_{\rm c}$.
In fact, due to successive discoveries of superconductivity
in strongly correlated electron systems such as
molecular conductors, transition metal oxides,
and heavy-fermion compounds,
it has been one of central issues in the research field
of condensed matter physics
to elucidate the mechanism of anisotropic superconductivity
with relatively high $T_{\rm c}$.
Among them, concerning a superconducting material group
characterized by singlet Cooper pair,
a key concept of $d$-wave superconductivity
mediated by antiferromagnetic spin fluctuations
has been believed to be established.
\cite{Miyake,review1,review2,review3,review3b,review4}

In addition to the concept of anisotropic superconductivity
mediated by magnetic fluctuations,
another important ingredient is multiband effect,
since superconductivity has been found in electron systems
with multiband such as Sr$_2$RuO$_4$,\cite{Maeno}
MgB$_2$,\cite{Akimitsu} and iron pnictides.\cite{Hosono,Ishida}
Since multi-sheets of Fermi surfaces have been usually
observed in heavy fermion compounds,
the occurrence of superconductivity in such materials
should be also related to multiband nature.
Just after the BCS theory,
multiband effect on $T_{\rm c}$ has been discussed
in a simple two-band electron model.\cite{twoband}
In fact, in recent years, multiband superconductivity
bas been actively investigated from various viewpoints.
\cite{Takimoto1,Bishop,Nomura,Takimoto2,Yanase,Yang,
Kontani,Dolgov,Bersier}
In particular, here we mention
$s_{\pm}$-wave superconductivity
proposed for iron pnictides.\cite{Mazin,Kuroki,Ikeda,Nomura2}

In general, in electron systems with degenerate orbitals,
Jahn-Teller phonons should play an important role,
since Jahn-Teller distortions are known to lift
the degeneracy in electron orbitals.
In fullerene superconductors,\cite{C60-1,C60-2,C60-3,C60-4}
$s$-wave pair formation due to Jahn-Teller phonons has
been discussed.
A possibility of superconductivity due to geometric phase
in Jahn-Teller crystals has been proposed.
\cite{Koizumi1,Koizumi2}
However, the attractive interaction mediated by Jahn-Teller phonons
has not been analyzed satisfactorily even in a weak-coupling limit,
probably because of tedious calculations to derive
such effective attractions in multiband systems.

In this paper, the effective interaction is derived
in the two-orbital electron system which is coupled with
breathing and Jahn-Teller phonons with the use of
a standard quantum mechanical techniques.
By applying a weak-coupling approximation to the effective model,
we discuss the enhancement of $T_{\rm c}$ due to Jahn-Teller phonons
in the two-band electron system.
When we include only the breathing phonon,
$T_{\rm c}$ is just the same as that of the famous BCS formula
for the one-band model.
On the other hand, when we consider Jahn-Teller phonons,
we find the significant enhancement of $T_{\rm c}$
by the interband attraction.
These weak-coupling solutions are checked from the numerical
estimation of pair susceptibility in the effective 4-site model.
It is concluded that $T_{\rm c}$ of the multiband systems
coupled with phonons is enhanced
by the number of relevant phonon modes.


Now we consider a two-orbital electron system
coupled with breathing and Jahn-Teller phonons,
given by
\begin{equation}
  \label{eq:H}
  \begin{split}
   H &= \sum_{\mib{i}\mib{a},\gamma\gamma'\sigma}
      t^{\mib{a}}_{\gamma\gamma'}d^{\dag}_{\mib{i}\gamma\sigma}
      d_{\mib{i}+\mib{a}\gamma'\sigma}
      + \sum_{\ell,\mib{i}} g_{\ell}Q_{\ell\mib{i}}\rho_{\ell\mib{i}} \\
     &+ \sum_{\ell,\mib{i}} [P_{\ell\mib{i}}^2/(2M_{\ell})
                            +k_{\ell} Q_{\ell\mib{i}}^2/2],
  \end{split}
\end{equation}
where $d_{\mib{i}\gamma\sigma}$ is an annihilation operator
for an electron with spin $\sigma$ in the orbital $\gamma$
(=$a$ and $b$) at site $\mib{i}$,
$t^{\mib{a}}_{\gamma\gamma'}$ denotes the electron hopping between
adjacent $\gamma$- and $\gamma'$-orbitals in nearest neighbor sites
connected by a vector $\mib{a}$,
$g_{\ell}$ indicates the coupling constant between electrons
and local distortion specified by $\ell$,
$Q_{1\mib{i}}$ indicates breathing distortion,
$Q_{2\mib{i}}$ and $Q_{3\mib{i}}$ denote Jahn-Teller distortions,
$\rho_{1\mib{i}}$=
$d_{{\mib i} a \sigma}^{\dag}d_{\mib{i} a \sigma}
+d_{{\mib i} b \sigma}^{\dag}d_{\mib{i} b \sigma}$,
$\rho_{2\mib{i}}$=
$d_{{\mib i} a \sigma}^{\dag}d_{\mib{i} b \sigma}
+d_{{\mib i} b \sigma}^{\dag}d_{\mib{i} a \sigma}$,
$\rho_{3\mib{i}}$=
$d_{{\mib i} a \sigma}^{\dag}d_{\mib{i} a \sigma}
-d_{{\mib i} b \sigma}^{\dag}d_{\mib{i} b \sigma}$,
$P_{\ell\mib{i}}$ denotes canonical momentum of $Q_{\ell\mib{i}}$,
$M_{\ell}$ is corresponding reduced mass,
and $k_{\ell}$ denotes the spring constant.

By following the standard procedure of quantization of phonons,
we introduce the phonon operator $a_{\ell\mib{i}}$, defined through
$Q_{\ell\mib{i}}$=
$(a_{\ell\mib{i}}+a_{\ell\mib{i}}^{\dag})/\sqrt{2\omega_{\ell} M_{\ell}}$,
where $\omega_{\ell}$ is the phonon energy given by
$\omega_{\ell}$=$\sqrt{k_{\ell}/M_{\ell}}$.
After performing the Fourier transform,
we obtain $H$ in the form of $H$=$H_0$+$H_1$,
where $H_0$ denotes the sum of electron and phonon energy,
given by
\begin{equation}
  \label{eq:H0}
  H_0 = \sum_{\mib{k},\tau\sigma}
        E_{\mib{k}\tau}
        d^{\dag}_{\mib{k}\tau\sigma}d_{\mib{k}\tau\sigma}
      + \sum_{\ell,\mib{q}} \omega_{\ell}
       (a_{\ell \mib{q}}^{\dag} a_{\ell \mib{q}}+1/2).
\end{equation}
Here the electron energy is given by
\begin{equation}
  \label{eq:Ek}
  E_{\mib{k}\tau}=\Bigl[
  \varepsilon_{\mib{k}aa} + \varepsilon_{\mib{k}bb} \pm
  \sqrt{(\varepsilon_{\mib{k}aa}-\varepsilon_{\mib{k}bb})^2
        +4\varepsilon_{\mib{k}ab}^2}
  \Bigr]/2,
\end{equation}
where $\tau$=1 and 2 correspond to $+$ and $-$ signs, respectively,
and $\varepsilon_{\mib{k}\gamma\gamma'}$=
$\sum_{\mib{a}}e^{i\mib{k} \cdot \mib{a}}t^{\mib a}_{\gamma\gamma'}$.
The electron-phonon coupling term $H_1$ is given by
\begin{equation}
  \label{eq:H1}
  H_1 \!=\! \sum_{\ell,\mib{k}\mib{q},\tau\tau'}\!
        \sqrt{\alpha_\ell} \omega_{\ell}
        (a_{\ell \mib{q}}+a_{\ell -\mib{q}}^{\dag})
        \Gamma^{(\ell)}_{\tau\tau'}(\mib{k},\mib{q})
        \rho_{\tau\tau'}(\mib{k},\mib{q}),
\end{equation}
where
$\alpha_{\ell}$=$g_{\ell}^2/(2M_{\ell} \omega_{\ell}^3)$,
$\rho_{\tau\tau'}(\mib{k},\mib{q})$=
$\sum_{\sigma}d^{\dag}_{\mib{k}+\mib{q}\tau\sigma}
d_{\mib{k}\tau'\sigma}$,
and the coefficient matrices ${\hat \Gamma}$'s are given by
\begin{equation}
 \begin{split}
  {\hat \Gamma}^{(1)}(\mib{k},\mib{q}) &=
  \left(
  \begin{array}{cc}
         A^+_{\mib{k},\mib{q}} & B^-_{\mib{k},\mib{q}} \\
        -B^-_{\mib{k},\mib{q}} & A^+_{\mib{k},\mib{q}} \\
  \end{array}
  \right),\\
  {\hat \Gamma}^{(2)}(\mib{k},\mib{q}) &=
  \left(
  \begin{array}{cc}
        -B^+_{\mib{k},\mib{q}} & A^-_{\mib{k},\mib{q}} \\
         A^-_{\mib{k},\mib{q}} & B^+_{\mib{k},\mib{q}} \\
  \end{array}
  \right),\\
  {\hat \Gamma}^{(3)}(\mib{k},\mib{q}) &=
  \left(
  \begin{array}{cc}
         A^-_{\mib{k},\mib{q}} & B^+_{\mib{k},\mib{q}} \\
         B^+_{\mib{k},\mib{q}} & -A^-_{\mib{k},\mib{q}} \\
  \end{array}
  \right).
\end{split}
\end{equation}
Here
$A^{\pm}_{\mib{k},\mib{q}}$=
$u^+_{\mib{k}+\mib{q}} u^+_{\mib{k}} \pm u^-_{\mib{k}+\mib{q}} u^-_{\mib{k}}$
and
$B^{\pm}_{\mib{k},\mib{q}}$=
$u^+_{\mib{k}+\mib{q}} u^-_{\mib{k}} \pm u^-_{\mib{k}+\mib{q}} u^+_{\mib{k}}$,
where $u^{\pm}_{\mib{k}}$ is given by
\begin{equation}
  u^{\pm}_{\mib{k}}=\sqrt{\frac{1}{2}} \Biggl[
  1 \pm \frac{\varepsilon_{\mib{k}aa}-\varepsilon_{\mib{k}bb}}
  {\sqrt{(\varepsilon_{\mib{k}aa}-\varepsilon_{\mib{k}bb})^2+4\varepsilon_{\mib{k}ab}^2}}
  \Biggr]^{1/2}.
\end{equation}
Since we assume degenerate the Jahn-Teller modes here,
we set $\omega_1$=$\omega_{\rm br}$ and
$\omega_2$=$\omega_3$=$\omega_{\rm JT}$.
Concerning coupling constants,
we introduce $\alpha_1$=$\alpha_{\rm br}$ and
$\alpha_2$=$\alpha_3$=$\alpha_{\rm JT}$.

Now we derive the effective Hamiltonian $H_{\rm eff}$ from $H$ due to the
elimination of phonon degrees of freedom by using a canonical transformation.
Let us here consider a transformation
$H_{\rm eff}$=$e^{S} H e^{-S}$,
where the operator $S$ is defined so as to satisfy the relation
$H_1$=$-[H_0, S]$.
Then, after some calculations of operators, we obtain
the effective Hamiltonian as $H_{\rm eff}$=$H_0+H_{\rm int}$.
The effective interaction between electrons
mediated by phonons is given by
\begin{equation}
  \label{eq:Hint}
  H_{\rm int}=-[[H_0,S],S]/2.
\end{equation}
Here we obtain the effective model within the second order of
electron-phonon coupling constant.

In the present case, first we assume $S$ in the form of
\begin{equation}
  \begin{split}
   S = \sum_{\ell,\mib{kq},\tau\tau'}
   & \sqrt{\alpha_\ell} \omega_\ell
     [X^{-(\ell)}_{\tau\tau'}(\mib{k},\mib{q})a_{\ell\mib{q}} \\
   & +X^{+(\ell)}_{\tau\tau'}(\mib{k},\mib{q})a_{\ell -\mib{q}}^{\dag}]
      \rho_{\tau\tau'}(\mib{k},\mib{q}),
  \end{split}
\end{equation}
where $X^{\pm(\ell)}_{\tau\tau'}$
is determined so as to satisfy $H_1$=$-[H_0, S]$.
After lengthy calculations, we obtain
\begin{equation}
  X^{\pm(\ell)}_{\tau\tau'}(\mib{k},\mib{q})=
  \frac{-\sqrt{\alpha_{\ell}}\omega_{\ell}}
  {E_{\mib{k}+\mib{q}\tau}-E_{\mib{k}\tau'} \pm \omega_{\ell}}
   \Gamma^{(\ell)}_{\tau\tau'}(\mib{k},{q}).
\end{equation}
The effective interaction $H_{\rm int}$ is evaluated by 
$[H_1,S]/2$ from eq.~(\ref{eq:Hint}).
Then, we obtain
\begin{equation}
\begin{split}
  H_{\rm int} =\sum_{\ell,\tau\tau',\mu\mu'}\sum_{\mib{k},\mib{k'},\mib{q}}
  & U_{\ell}\frac{\omega_{\ell}^2
  \Gamma^{(\ell)}_{\tau\tau'}(\mib{k},\mib{q})
  \Gamma^{(\ell)}_{\mu\mu'}(\mib{k}',-\mib{q})}
  {(E_{\mib{k}'\tau'}-E_{\mib{k'}-\mib{q}\tau})^2-\omega_{\ell}^2} \\
  &\times \rho_{\tau\tau'}(\mib{k},\mib{q})
          \rho_{\mu\mu'}(\mib{k}',-\mib{q}),
\end{split}
\end{equation}
where $U_{\ell}$=$\alpha_{\ell}\omega_{\ell}$
and $U_{\ell}$ denotes
the attractive interaction due to $\ell$-mode phonons.
Note here that $U_1$=$U_{\rm br}$=$\alpha_{\rm br}\omega_{\rm br}$
and $U_2$=$U_3$=$U_{\rm JT}$=$\alpha_{\rm JT}\omega_{\rm JT}$.

\begin{figure}[t]
\label{fig1}
\centering
\includegraphics[width=7.5truecm]{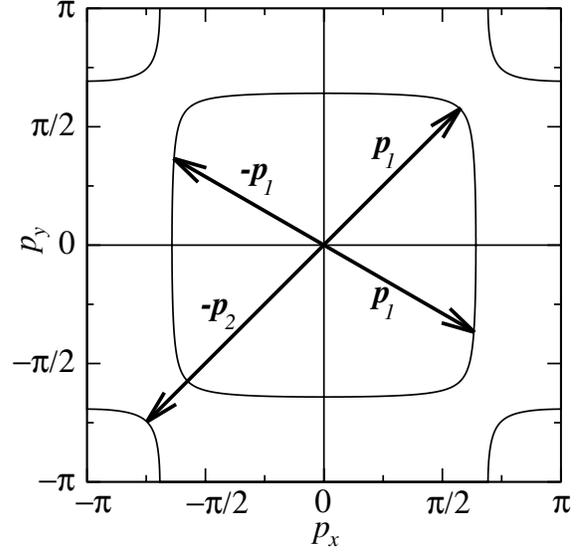}
\caption{
Cooper pairs in a two-dimensional electron system
with a couple of Fermi surfaces.
In the weak-coupling limit, pairs are formed by electrons
on the same Fermi surface, since the pair between different
Fermi surfaces should be composed of electrons with
$\mib{p}_1$ and $-\mib{p}_2$.
}
\end{figure}

Here we provide a comment on the Cooper pair in the system
with multi-Fermi surfaces.
In the weak-coupling limit, we usually consider the pairing
of electrons only in the vicinity of the Fermi energy $E_{\rm F}$.
Thus, we set
$E_{\mib{k'}-\mib{q}\tau}$=$E_{\mib{k'}\tau'}$$\approx$$E_{\rm F}$.
As shown in Fig.~1, in the weak-coupling limit,
the Cooper pair is formed only by the electrons
on the same Fermi surface,
except for an unrealistic case in which a couple of Fermi-surface
sheets are perfectly degenerate.
Then, we consider the interaction for Cooper pair
with zero total momentum.
Note that in the strong-coupling region, it is possible to consider
the pairs of electrons far from the Fermi surface,
leading to a chance of pair formation
between different Fermi surfaces.

After some algebraic calculations,
we obtain the effective interaction in the form of
\begin{equation}
  H_{\rm int} = -\sum_{\mib{p},\mib{p}',\mu\nu}
  V_{\mu\nu}(\mib{p},\mib{p}')
  d^{\dag}_{\mib{p}\mu\uparrow}d^{\dag}_{-\mib{p}\mu\downarrow}
  d_{-\mib{p}'\nu\downarrow}d_{\mib{p}'\nu\uparrow}.
\end{equation}
The pair potential $V_{\mu\nu}$ is given by
\begin{equation}
  \begin{split}
   & V_{11}(\mib{p},\mib{p}') = V_{22}(\mib{p},\mib{p}')
     =U_{\rm JT}+U_{\rm br} A^{+2}_{\mib{p}',\mib{p}-\mib{p}'},\\
   & V_{12}(\mib{p},\mib{p}') = V_{21}(\mib{p},\mib{p}')
     =U_{\rm JT}+U_{\rm br} B^{-2}_{\mib{p}',\mib{p}-\mib{p}'}.
  \end{split}
\end{equation}
Here we note the relation of
$A^{+2}_{\mib{p}',\mib{p}-\mib{p}'}$+
$B^{-2}_{\mib{p}',\mib{p}-\mib{p}'}$=1.

Next we solve the gap equation.
The gap function $\Delta_{\mu}(\mib{p})$ is given by
\begin{equation}
   \Delta_{\mu}(\mib{p}) = \sum_{\mib{p}',\nu}
   V_{\mu\nu}(\mib{p},\mib{p}')
   \langle
   d_{-\mib{p}'\nu\downarrow}d_{\mib{p}'\nu\uparrow}
   \rangle,
\end{equation}
where $\langle \cdots \rangle$ denotes the average
by using $H$ in the mean-field approximation.
Note again that it is enough to consider the pairs on
the same Fermi surfaces in the weak-coupling limit.
By assuming the Cooper pair with $s$-wave symmetry,
we obtain the gap equation at $T$=$T_{\rm c}$ as
\begin{equation}
   \label{eq:gap}
   \Delta_{\mu} = \log (1.13 \omega_{\rm c}/T_{\rm c}) \sum_{\nu}
   \lambda_{\mu \nu} \Delta_{\nu},
\end{equation}
where $\omega_{\rm c}$ is an appropriate cut-off frequency
and $\lambda_{\mu \nu}$ is the non-dimensional coupling constant,
given by
\begin{equation}
  \label{eq:int}
  \begin{split}
   & \lambda_{11} = \lambda_{22}
   = \lambda_{\rm JT}+ \lambda_{\rm br} \beta,\\
   & \lambda_{12} = \lambda_{21}
   = \lambda_{\rm JT}+ \lambda_{\rm br} (1-\beta).
  \end{split}
\end{equation}
Here $\lambda_{\rm JT}$=$N_0 U_{\rm JT}$,
$\lambda_{\rm br}$=$N_0 U_{\rm br}$,
$N_0$ denotes the density of states at the Fermi level,
$\beta$=$\langle A^{+2}_{\mib{p}',\mib{p}-\mib{p}'} \rangle_{\rm FS}$,
and $\langle \cdots \rangle_{\rm FS}$ denotes
the average over the Fermi surface.
Note that we simply assume the same values of $N_0$
for the different Fermi surfaces.

By solving the gap equation eq.~(\ref{eq:gap}),
we obtain
\begin{equation}
   \label{eq:Tc}
   T_{\rm c}
   = 1.13 \omega_{\rm c}e^{-1/(\lambda_{11} + \lambda_{12})}
   = 1.13 \omega_{\rm c}e^{-1/(2\lambda_{\rm JT} + \lambda_{\rm br})}.
\end{equation}
Note that another solution provides smaller $T_{\rm c}$ even if it exists.
Equation (\ref{eq:Tc}) tells us several interesting stories.
First we consider a situation in which only the breathing mode is active.
In this case, we immediately obtain the same formula of $T_{\rm c}$,
$T_{\rm c}$=$1.13 \omega_{\rm c} e^{-1/\lambda_{\rm br}}$,
as that of the BCS theory for a one-band system.
Namely, even if the number of the band is increased,
the magnitude of $T_{\rm c}$ is not changed 
as long as the total electron density is coupled with
the breathing phonons.
In this situation, there is no advantageous points of multi-band nature
for the elevation of $T_{\rm c}$.

\begin{figure}[t]
\label{fig2}
\centering
\includegraphics[width=7.5truecm]{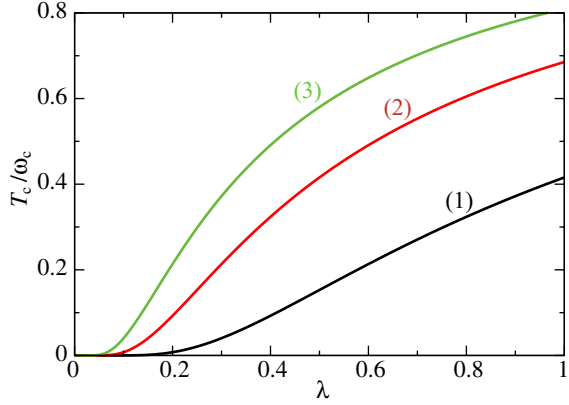}
\caption{
(Color online)
Curves of
(1) $T_{\rm c}/\omega_{\rm c}=1.13e^{-1/\lambda_{\rm br}}$,
(2) $T_{\rm c}/\omega_{\rm c}=1.13e^{-1/(2\lambda_{\rm JT})}$,
and
(3) $T_{\rm c}/\omega_{\rm c}=1.13e^{-1/(2\lambda_{\rm JT}+\lambda_{\rm br})}$.
Here we set $\lambda_{\rm JT}=\lambda_{\rm br}=\lambda$
for simplicity.
}
\end{figure}

On the other hand, in the multi-orbital system,
there occurs a coupling with Jahn-Teller phonons
so as to lift the degeneracy in electron systems.
In such a case, the factor 2 appears in front of the coupling constant
in the $T_{\rm c}$ formula.
In other words, this factor 2 indicates the number of Jahn-Teller modes,
{\it not} the number of electron bands.
When we consider the coupling of degenerate electrons with
both breathing and Jahn-Teller phonons,
the factor 3 becomes effective if we simply consider
$\lambda_{\rm br}$=$\lambda_{\rm JT}$.
In Fig.~2, we show the curves of $T_{\rm c}$
for the three cases of
(1) breathing phonon, (2) Jahn-Teller phonons,
and (3) both breathing and Jahn-Teller phonons.
Here for simplicity,
we set $\lambda_{\rm br}$=$\lambda_{\rm JT}$=$\lambda$.
As easily understood, the change of the factor
in the power is remarkable,
even in the weak-coupling approximation.
Note that the factor 2 or 3 indicates the total number
of phonon modes which are coupled with electron systems.

In order to confirm the present result of
eq.~(\ref{eq:Tc})
obtained in the weak-coupling BCS approximation,
we evaluate the pair susceptibility of
the effective two-band model with the attractive
interaction induced by phonons.
For the purpose, we resort to an unbiased technique
such as exact diagonalization.
The singlet pair correlation function
is evaluated in a small-sized cluster
and the effective coupling constant $\lambda_{\rm eff}$
is deduced from the singlet pair correlation.

The effective model is given by
\begin{equation}
 \label{eq:Heff}
 \begin{split}
  H_{\rm eff} &= \sum_{\mib{k},\tau\sigma}
       E_{\mib{k}\tau} d^{\dag}_{\mib{k}\tau\sigma} d_{\mib{k}\tau\sigma}
   - I \sum_{\mib{i},\tau}
  d^{\dag}_{\mib{i}\tau\uparrow} d_{\mib{i}\tau\uparrow}
      d^{\dag}_{\mib{i}\tau\downarrow} d_{\mib{i}\tau\downarrow} \\
  &- J \sum_{\mib{i}}
   (d^{\dag}_{\mib{i}1\uparrow} d^{\dag}_{\mib{i}1\downarrow}
      d_{\mib{i}2\downarrow}d_{\mib{i}2\uparrow}+{\rm h.c.}),
\end{split}
\end{equation}
where $I$ and $J$ denote on-site and pair-hopping {\it attractive}
interactions corresponding to $V_{11}$ and $V_{12}$, respectively.
In order to reproduce the situation in eq.~(\ref{eq:int}),
we set $I$ and $J$ as
$I=U_{\rm JT} + U_{\rm br} \beta$
and
$J=U_{\rm JT} + U_{\rm br} (1-\beta)$,
respectively, by taking $\beta$ as a parameter.

The pair susceptibility matrix $\chi_{\mu\nu}({\mib{m},\mib{n}})$
is defined by
\begin{equation}
  \chi_{\mu\nu}(\mib{m},\mib{n})=\int_0^{1/T} d\tau
  \langle {\hat \phi}_{\mib{m},\mu}(\tau)
  {\hat \phi}^{\dag}_{\mib{n},\nu} \rangle,
\end{equation}
where $T$ is a temperature,
${\hat \phi}_{\mib{m},\mu}(\tau)$=
$e^{H\tau}{\hat \phi}_{\mib{m},\mu} e^{-H\tau}$,
$\mib{m}$ indicates the vector connecting
possible two sites in the cluster,
and ${\hat \phi}_{\mib{m},\mu}$ is a singlet pair operator
of the band $\mu$, given by
\begin{equation}
  {\hat \phi}_{\mib{m},\mu}=\sum_{\mib{i}}
  \phi_{\mib{m},\mu}
  (d_{\mib{i}\mu\downarrow}d_{\mib{i}+\mib{m}\mu\uparrow}
  -d_{\mib{i}\mu\uparrow}d_{\mib{i}+\mib{m}\mu\downarrow})\sqrt{2}.
\end{equation}
The coefficient $\phi_{\mib{m},\mu}$ is determined by
the diagonalization of the susceptibility matrix
and the pair susceptibility $\chi_{\mu\nu}$ is defined 
by its maximum eigenvalue.

In order to extract information on the pairing interaction,
we consider the pair susceptibility in a diagrammatic manner.
When we define the non-interacting pair susceptibility
as ${\hat \chi}^{(0)}$,
${\hat \chi}$ satisfies the relation of
${\hat \chi}$=${\hat \chi}^{(0)}$+
${\hat \chi}^{(0)}{\hat K}{\hat \chi}$
in the ladder approximation,
where ${\hat K}$ denotes the effective pairing interaction
in the matrix form.
Usually we obtain ${\hat \chi}$ from ${\hat K}$
and ${\hat \chi}^{(0)}$,
but here we evaluate ${\hat K}$ as
${\hat K}$=${\hat \chi}^{(0)-1}-{\hat \chi}^{-1}$,
where ${\hat \chi}$ is numerically evaluated.
Then, we can evaluate the effective
non-dimensional coupling constant $\lambda_{\rm eff}$ as
\begin{equation}
 \lambda_{\rm eff}=(K_{11}+K_{12})/W,
\end{equation}
where $W$ is the bandwidth.

In this paper, for the evaluation of ${\hat \chi}$,
we exploit an exact diagonalization technique
for the model in a 4-site cluster.
We set $E_{\mib{k}1}=-t(\cos k_x + \cos k_y)$
and $E_{\mib{k}2}=t(\cos k_x + \cos k_y)$,
leading to a couple of Fermi surfaces around
$\Gamma$ and $M$ points in the thermodynamic limit.
Here we consider the case of $n$=0.5,
where $n$ is the electron number per site and per orbital.
In this case,
$\chi^{(0)}_{11}$=$\chi^{(0)}_{22}$=$4/t$
and $W$=$4t$ in the 4-site cluster.

\begin{figure}[t]
\label{fig3}
\centering
\includegraphics[width=8.0truecm]{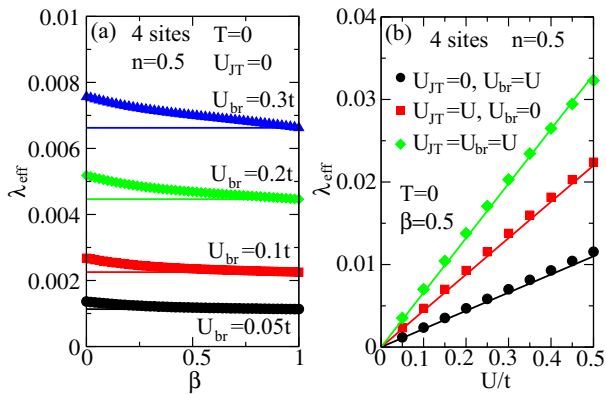}
\caption{
(Color online) (a) Effective coupling constant $\lambda_{\rm eff}$
vs. $\beta$ for $U_{\rm br}/t$=$0.05$, $0.1$, $0.2$, and $0.3$
with $n$=$0.5$, $U_{\rm JT}$=0, and $T$=0.
As for the horizontal lines, see the main text.
(b) $\lambda_{\rm eff}$ vs. $U/t$
for ($U_{\rm br}$, $U_{\rm JT}$)=($U$,0),
(0,$U$) and ($U$,$U$) with $\beta$=0.5.
Three lines indicate $\lambda_{\rm eff}$
=$0.022\times U/t$, $0.044 \times U/t$,
and $0.066 \times U/t$.
}

\end{figure}

In Fig.~3(a), we show $\lambda_{\rm eff}$ vs. $\beta$
by solid symbols
for $U_{\rm br}$=$0.05t$, $0.1t$, $0.2t$, and $0.3t$
with $n$=0.5, $U_{\rm JT}$=0, and $T$=0.
Note that each horizontal line denotes the pair susceptibility
of a one-band model with nearest neighbor hopping $t$
and on-site attraction $U_{\rm br}$
for the same value of $n$ in the 4-site cluster.
For small values of $U_{\rm br}/t$,
we find that $\lambda_{\rm eff}$ weakly depends on $\beta$
and it agrees well with the result of the one-band model,
suggesting that $T_{\rm c}$ in the two-band system
coupled with breathing phonons is just the same as
that of the one-band model.
Namely, in such a case, $T_{\rm c}$ is not expected to
increase even if we increase the number of electron bands.

In Fig.~3(b), we show $\lambda_{\rm eff}$ vs. $U/t$
with a fixed value of $\beta$=$0.5$ for three cases as
(1) $U_{\rm br}$=$U$ and $U_{\rm JT}$=$0$,
(2) $U_{\rm br}$=$0$ and $U_{\rm JT}$=$U$,
and (3) $U_{\rm br}$=$U_{\rm JT}$=$U$.
The effective coupling constant is found to be in proportion
to $U/t$.
Among the proportional coefficients, we confirm the relation of
$\lambda_{\rm eff}^{(1)}$=
$\lambda_{\rm eff}^{(2)}/2$=
$\lambda_{\rm eff}^{(3)}/3$,
which is consistent with the weak-coupling result of
eq.~(\ref{eq:Tc}).


Thus far we have analyzed the models in the weak-coupling region,
but we are also interested in the strong-coupling behavior.
For instance, when we increase $U_{\rm br}$ in Fig.~3(a),
$\beta$-dependence of $\lambda_{\rm eff}$ becomes more significant
and the deviation from the one-band result is large.
When we increase the value of attractive interaction in Fig.~3(b),
$\lambda_{\rm eff}$ is no longer in proportion to $U/t$
and the relation of $\lambda_{\rm eff}^{(1)}$=
$\lambda_{\rm eff}^{(2)}/2$=
$\lambda_{\rm eff}^{(3)}/3$ does not hold.
In order to discuss such strong-coupling effects,
we should analyze directly the original
Hamiltonian eq.~(\ref{eq:H}), not the effective model
eq.~(\ref{eq:Heff}), by applying the Migdal-Eliashberg theory.
It is one of future problems.

Here we provide a brief comment on the polaron mass enhancement,
which is one of strong coupling effects.
For Holstein phonons, it has been well known that
the polaron mass $m^*$ is increased as
$m^*/m$=$e^{\alpha_{\rm br}}$,
where $m$ is the bare electron mass,
while for Jahn-Teller phonons,
the mass enhancement has been found to be expressed by
$m^*/m$$\approx$$e^{\alpha_{\rm JT}/2}/\sqrt{\alpha_{\rm JT}}$
for large $\alpha_{\rm JT}$.
Namely, for the same value of the coupling constant,
the mass of the Jahn-Teller polaron
is smaller than that of the Holstein polaron.
This behavior seems to be related to the fact that
the vertex corrections in an electron system coupled
with Jahn-Teller phonons
should be less effective in comparison
with the case of Holstein phonons.\cite{Takada1,Takada2}
The fact may be also relevant to the increase of superconducting
$T_{\rm c}$ in electron systems coupled with Jahn-Teller phonons.

Note also that in the Migdal-Eliashberg theory,
the effect of Coulomb interaction is included
in the parameter $\mu^*$ as the reduced repulsion.
The same story is expected to be applied to the present case,
as long as we consider adiabatic phonons.
However, the degree of the reduction of Coulomb repulsion
may be different between intra- and inter-orbital interactions.
If the on-site interaction is still negative
while the pair hopping interaction becomes positive,
we obtain the so-called $s_{\pm}$-wave pairing,
proposed for iron pnictides.
\cite{Mazin,Kuroki,Ikeda,Nomura2}
Thus, it may be possible to construct
an alternative Jahn-Teller phononic scenario
for superconductivity in iron pnictides.
In particular, competition between Coulomb repulsion
and phonon-induced attraction may be a key issue to
understand the appearance of both nodal $d$-wave
and nodeless $s_{\pm}$-wave gaps in iron pnictides.
This is an interesting problem in future.


In summary, we have discussed the appearance of
superconductivity in the two-band electron system
coupled with breathing and Jahn-Teller phonons
in the weak-coupling limit.
It has been found that $T_{\rm c}$ is increased
with the increase of the number of relevant phonon modes.
Namely, $T_{\rm c}$ of the two-band system coupled with
Jahn-Teller phonons becomes high in comparison with
that of the one-band case, leading to a possibility
of phonon-induced high-$T_{\rm c}$ superconductivity.


This work has been supported by a Grant-in-Aid
for Scientific Research on Innovative Areas ``Heavy Electrons''
(No. 20102008) of The Ministry of Education, Culture, Sports,
Science, and Technology, Japan.


\end{document}